\documentclass[nofootinbib,onecolumn,preprintnumbers,amsmath,amssymb,prd]{revtex4}

\preprint{IPM/P-2008/054}
\allowdisplaybreaks

\begin{document}
\title{Lower bound on the mass of a black hole}

\author{Qasem \surname{Exirifard}}
\affiliation{School of Physics, Institute for Studies in Theoretical Physics and Mathematics (IPM), P.O.Box 19395-5531, Tehran, Iran}

\email{exir@theory.ipm.ac.ir, exirifard@gmail.com}

\begin{abstract}
We consider gravity coupled to a massive field whose Compton's wavelength is far larger than the Planck's length. In the low energy effective action for gravity, thus, it is the perturbation in the Compton's wavelength that breaks first as the sub-sub-leading quantum perturbation grows stronger.  When this break occurs, we  can not trust  the perturbative information about the form of the low energy effective action. We translate this break into the lowest limit on the mass of a classical black hole. In $D=4$, using the electron's mass, this requires the black hole to be heavier than $10^{14} kg$.  
\end{abstract}

\maketitle

Pure gravity includes just one length scale: the Planck's length. Its low energy effective action around flat space-time, therefore, is a perturbative series like;
\begin{equation}
L_{eff} \,=\, R + l_p^2 (c_1 R^2 + c_2 R_{\mu\nu} R^{\mu\nu} + c_3 R_{\mu\nu\lambda\eta} R^{\mu\nu\lambda\eta})+\cdots\,, 
\end{equation}
where $c_i$'s are constant numbers at order one and $l_p$ stands for the (reduced) Planck's length:
\begin{eqnarray}
m_{p} & =& \sqrt{\frac{\hbar c}{8 \pi G}} \,=\, 4.34 \mu g\,,\\
l_p & =& \frac{\hbar}{c m_p} \,. 
\end{eqnarray}
The low energy effective action for pure gravity, thus, can be re-expressed in the following compact form
\begin{eqnarray}
L \,=\, R + \sum l_p^{2n} L^{(n)} (R_{\mu\nu\lambda\eta},\nabla_\mu,g_{\mu\nu})\,, 
\end{eqnarray}
where $L^{(n)}$ is a scalar with length dimension of $[L^{(n)}]=-2-2n$ constructed out from the Riemann tensor and its covariant derivatives.

Nature, however, has occurred not to be pure gravity. Its Platonic realization should also at-least include the content of standard model of particle physics. The mass scale of standard model is extra-ordinarily smaller than that of gravity. For example the Compton wavelength of electron is about $10^{22}$ times larger than the Planck's length. Approximating nature to pure gravity model that includes just one length scale, thus, sounds not physical in my point of view. So, what does the low energy effective action of (physical) gravity look like? This note aims to reply this question.   

Let a field of mass $m$ ($m\ll m_p$) couple to gravity. The low energy effective action for metric couple to this field has two length scales, the Planck's length and the Compton wavelength of the massive particle: $l_p, l_c$. The low energy effective action for the metric reads
\begin{equation}
 L \,=\, L(R_{\mu\nu\lambda\eta,\nabla_\mu},g_{\mu\nu},l_p,l_c)
\end{equation}
When we expand the low energy effective action around the flat space-time, we get a double expansion series:
\begin{eqnarray}
L &=& R + l_p^2\, (L^{(1,0)} \,+\, l_c^2 L^{(1,1)} \,+\, l_c^4 L^{(1,2)} \,+\, \cdots) \\
&& ~+  l_p^4 \,(L^{(2,0)} \,+\, l_c^2 L^{(2,1)} \,+\, l_c^4 L^{(2,2)} \,+\, \cdots) \nonumber \\
&& ~+\cdots \nonumber 
\end{eqnarray}
where $L^{(m,n)}$ is a scalar with appropriate dimension constructed out from the Riemann tensor and its covariant derivatives. Also notice that in the above series, I have excluded terms that carry odd power of the Planck's length.\footnote{Also note that the limit of $m \to 0$ may not coincide to $m=0$. The quantum corrections in ordinary QED do not vanish when $m_e\to 0$, for example look at \cite{Davila:2009vt}. The sub-leading corrections in QED will influence the sub-sub-leading corrections to effective action of metric. So there exists no reason that the limit of $m_e\to 0$ of the low energy effective action for the metric is $m_e=0$.} This exlusion is due to the fact that $l_p\propto \sqrt{\hbar}$ beside observing that in QFT all perturbations are at order of $\hbar$ power to a natural number. 

Let's look at the double expansion series. As the curvature of the space-time increases, the perturbative terms grow stronger. Since $l_c >>l_p$, it is the perturbation in $l_c$ that breaks first. From the moment that this break occurs, we can not trust the perturbative series anymore. Physically (loosely) speaking when the curvature of the space-time becomes comparable to the Compton's wavelength of the particle then quantum fluctuations of the space-time can create a large number of the particle, such a large number that gravitational back reaction of the produced particle can not be necessarily neglected. 

When the scale of the curvature of the space-time is at the order or smaller than the Compton's wavelength of the particle, we should first calculate all the perturbative sub-corrections in $l_c$, and next sum the series before we address what quantum effects are. Summing all the $l_c$-terms leads to 
\begin{equation}
L = R + \sum_{n=0}^{\infty} \frac{l_p^{2n}}{l_c^{2n+2}} L^{(n)}(l_c^2 R_{\mu\nu\lambda\eta},l_c \nabla_\mu, g_{\mu\nu}) 
\end{equation}
 where $L^{(n)}(l_c^2 R_{\mu\nu\lambda\eta},l_c \nabla_\mu, g_{\mu\nu})$ is an scalar, not necessarily a polynomial, constructed out from its arguments. It is an interesting observation that $\frac{l_p^2}{l_\nu^4}$ where $l_\nu$ is the Compton wavelength associated to the lightest massive neutrino is almost at the order of the observed cosmological constant. So far, no one has succeeded to calculate all the $l_c$ terms in a consistent theory (if we have any) and sum all these terms together to find  the exact form of $L^{(n)}(l_c^2 R_{\mu\nu\lambda\eta},l_c \nabla_\mu, g_{\mu\nu})$.

Let a Schwarzschild black hole in $D=4$ be considered
\begin{equation}
ds^2 \, =\, - (1- \frac{2 G M_{BH}}{c^2 r}) dt^2 + \frac{dr^2 }{1- \frac{2 G M_{BH}}{c^2 r}} + r^2 d\Omega^2\,. 
\end{equation}
The lower upper bound on the Riemann scalar curvature outside the horizon of the black hole reads
\begin{equation}\label{above}
Max (R_{\mu\nu\lambda\eta} R^{\mu\nu\lambda\eta}) \propto (\frac{m_p}{M_{BH}})^4 \frac{1}{l_p^4}\,. 
\end{equation}
when $R_{\mu\nu\lambda\eta} R^{\mu\nu\lambda\eta} \propto \frac{1}{l_c^4}$ then the perturbative series in $l_c$ breaks. Using \ref{above}, therefore, gives a lower limit on the mass of a black hole. This lower limit turns to be
\begin{equation}
 M_{BH} > \frac{m_p^2}{m}
\end{equation}
where $m$ is the mass of the massive particle. Considering the electron and the ordinary Planck's mass, we obtain that the smallest black hole should be heavier than  $10^{14} kg$. If we consider extra dimensional scenario to bring down the fundamental scale of gravity to TeV or so, the smallest black hole will be much heavier than TeV. The lower limit of mass of a Schwarzschild black hole in a $D$ dimensional space-time reads
\begin{equation}
 M_{BH} > M_p (\frac{M_p}{m})^{D-3}
\end{equation}
where $M_p$ is the fundamental scale of gravity. If we brings down the scale of gravity to $1 TeV$ in a $D=5$, then using the electron's mass, we see that the lowest black hole should be heavier than $10^{12} TeV$. So no human-made accelerator can produce what we could be sure that is a black hole, in no higher dimensional scenario. 

\providecommand{\href}[2]{#2}\begingroup\raggedright


\begin{thebibliography}{10}
%\cite{Davila:2009vt}
\bibitem{Davila:2009vt}
  J.~M.~Davila and C.~Schubert,
  \textit{Effective action for Einstein-Maxwell theory at order RF**4,}
  arXiv:0912.2384 [gr-qc].
  %%CITATION = ARXIV:0912.2384;%%


 \end{thebibliography}
\end{document}